\documentclass{article}

\usepackage{arxiv}

\usepackage[utf8]{inputenc} 
\usepackage[T1]{fontenc}    
\usepackage{hyperref}       
\usepackage{url}            
\usepackage{booktabs}       
\usepackage{amsfonts}       
\usepackage{nicefrac}       
\usepackage{microtype}      
\usepackage{lipsum}		
\usepackage{graphicx}

\usepackage{csquotes}

\title{Resonating Experiences of Self and Others enabled by a Tangible Somaesthetic Design}

\author{{Ilhan Aslan, Andreas Seiderer, Chi Tai Dang, Simon Rädler, and Elisabeth André}\\
Human-Centered Multimedia Lab\\
Augsburg University\\
Germany\\
\texttt{lastname@hcm-lab.de} 
}

\begin{document}
\maketitle

\begin{abstract}
Digitalization is penetrating every aspect of everyday life including a human's heart beating, which can easily be sensed by wearable sensors and displayed for others to see, feel, and potentially ``bodily resonate'' with.
Previous work in studying human interactions and interaction designs with physiological data, such as a heart's pulse rate, have argued  that feeding it back to the users may, for example support users' mindfulness and self-awareness during various everyday activities and ultimately support their wellbeing. 
Inspired by Somaesthetics as a discipline, which focuses on an appreciation of the living body's role in all our experiences, we designed and explored mobile tangible heart beat displays, which enable rich forms of bodily experiencing oneself and others in social proximity.
In this paper, we first report on the design process of tangible heart displays and then present results of a field study with 30 pairs of participants.  Participants were asked to use the tangible heart displays during watching movies together and report their experience in three different heart display conditions (i.e., displaying their own heart beat, their partner's heart beat, and watching a movie without a heart display). We found, for example that participants reported significant effects in experiencing sensory immersion when  they felt their own heart beats compared to the condition without any heart beat display, and that feeling their partner's heart beats resulted in significant effects on social experience. We refer to resonance theory to discuss the results, highlighting the potential of how ubiquitous technology could utilize physiological data to provide resonance in a modern society facing social acceleration.    		
	
\end{abstract}


\section{Introduction}

A healthy human body is arguably an intelligent, self-regulating, and self-maintaining organism, which requires little explicit attention to function and enable joyful everyday experiences. However, as a consequence of unhealthy behavior, stress, age, injuries and illnesses of various sorts, etc. our bodies will need attention to ensure quality of life. Technology and digitalization may assist users in ubiquitously observing their bodily behavior and potentially help to correct any misbehavior \cite{Shusterman2008}.   

Since digitalization is in the process of transforming the health domain, it is not surprising that the body of related work is increasing rapidly in terms of the diversity of explored applications, user experiences, technologies, and interaction designs.  
For example, Dang et al.\ \cite{Dang:2019:TSS} have recently proposed to use an affective mirror in a smarthome setting to feed back a machine interpretation of a user's expressed emotion in addition to mirroring their physical image. They argue, for example, that feeding back to users how their affective states could be  interpreted by the ``outside world'', allows users to become aware of their emotional expressions and potentially optimize their self-presentation. To date, a multitude of data display technologies and techniques have been studied to inform and educate users about their own behavioral patterns and physiological and mental states, including tangible displays to feed back heart beats or breathing patterns to users through animating physical artifacts. For example,  Aslan et al.\ \cite{Aslan2016} have argued that a breathing plush toy or a beating artificial physical heart provide more natural and direct mappings of real heart beats and breathing movements than screen-based solutions. A part of their exploration was an inquiry with experts in mindfulness based stress reduction (MBSR) therapy \cite{Kabat2003}. Those experts had suggested that the main benefit of utilizing  mobile and multimodal displays could be in addressing the growing diversity in users and in application contexts (e.g., children in schools or inmates in prisons), enabling different user groups ubiquitous access to mindfulness practices and to their benefits, such as enhanced wellbeing \cite{Nyk2008} or stress reduction \cite{Shapiro2005}. 

Mobile technology, especially smartphones and apps have already started to provide easier and ubiquitous access to (traditionally analog) mindfulness practices, which have only been available to a group of dedicated and skilled individuals and meditation professionals who often practiced in environments and spaces specifically designed for meditation and mindfulness training. The surrounding world has undoubtedly an influence on users' experiences and their cognitive processes. For example, distributed cognition is a well known field and phenomena addressed in mobile and ubiquitous computing research \cite{Abowd2002}, and biofeedback loops are increasingly used to design for affective loop experiences \cite{hook2016move}.

Hartmut Rosa \cite{rosa2019resonance} who has recently introduced ``resonance theory'' to explain our relationships to the world argues that a modern ``accelerating'' world (e.g., a world in which we meet more and more people and own things for shorter times) increasingly fosters experiences of alienation and that people strive towards (ubiquitous experiences of) resonance. Hartmut Rosa argues that the increase of interest in yoga, wellbeing and mindfulness application, etc.  is an effect of social acceleration and people looking for resonance. 
Arguably, experiences of resonance can be achieved by assisting users in strengthening their relationships to people and potentially themselves.  We believe that an artificial physical heart that looks like and beats like a real heart has the potential to enable such resonating experiences by either feeding back users their own heartbeat or feeding them back a person's heartbeat to whom they relate to.

In order to explore how a tangible heart display influences users' experiences (especially experiences about how users relate to their surrounding world, such as immersion and empathy), we conducted a series of inquiries, including a field study with 30 pairs of users (i.e., 60 users). The field study utilized two tangible heart displays and focused on participants' ``in-movie''  and social experiences in three different conditions of watching together movies,  (i) while holding their own heart display, (ii) holding their partner's heart display, (iii) and using no heart display at all. We found, for example, a significant effect of feeling one's partner's heart beats on empathy and a significant effect of feeling one's own heart beats on immersion. 

Before we present methodology and discuss all the results of the field study in detail, we first provide background including related work, and describe the design and implementation of the tangible heart displays.

\section{Background}
Literature related to our work is spread over two well distinguishable research areas, that is (1) tangible heart rate interfaces and according user experiences gained in social contexts and (2) mindfulness and psychological well-being.
In the following, we address both topics with the most related works and show their relevance in order to lay a background for the remaining paper.

\subsection{Tangible Heart Rate Interfaces and User Experience in Social Contexts}
In UbiComp literature, there are numerous works devoted to the topic of tangible heart rate interfaces, ranging from novel methods for measuring heart rates (e.g., \cite{Chigira:2014:HRMonitor, Griffiths:2014:HRMonitor, Costa:2016:EmotionCheck}) to studies on user experience with such interfaces (e.g., \cite{Werner:2008:HeartRings, Hoinkis:2012:Herzfassen, Woodward:2018:EmoEcho, Howell:2019:HeartBench}).
The most widespread method of measuring heart rate today is based on an optical measurement technique using a wearable on the wrist.
In a recent work from UbiComp literature, for example, Costa et al. designed a wearable device called EmotionCheck \cite{Costa:2016:EmotionCheck} that measured a user's heart rate on user's wrist and generated feedback through tactile vibration patterns on the underside of the wearable.
They successfully showed that user's emotions, i.e., anxiety, can be manipulated through false feedback. In particular, they could regulate a user's anxiety in case of a slow heart rate.
Beyond that, in recent UbiComp works, researchers have been looking into ways to determine heart rate using everyday objects from the environment.
For example, Chigira et al. \cite{Chigira:2014:HRMonitor} utilized a photo-based HR sensor and the surface of transparent glass to measure heart rate during beverage consumption.
A chair as everyday object also provides a good opportunity to measure heart rate, as demonstrated by Griffiths et al. in ``Health Chair'' \cite{Griffiths:2014:HRMonitor}. 
For this purpose, they instrumented the armrests and backrest of a standard chair with the electrodes of an ECG to determine the heart rate.
Within the evaluation, the authors showed that the heart rate could be measured in 38\% of the time with an accuracy of 83\%, since the sitting position of a user and the position of the arms contribute significantly to whether a heart rate can be measured or not.
In contrast, our work goes beyond measuring approaches and explores the user experiences of tangible heart rate interfaces in a social context.

The research on user experiences with tangible interfaces that incorporate heart rate can be assigned to one or more of the categories games, art, awareness, and shared or remotely shared experiences. 
In the following, we briefly discuss an overview of the works most relevant to our paper.
Nearly all works address the (self-) awareness, which is why we discuss this aspect in the corresponding works, in case it is of importance.

In terms of tangible games, the heart rate is mostly used straightforwardly within the interface, for example to inform a player about his nervousness as demonstrated by Dang et al. in ``Surface-Poker'' \cite{Dang:2010:Poker}.
A more challenging approach is to adjust or influence game parameters based on the heart rate. For example, in ``Heartbeat Jenga'' presented by Huang et al. \cite{Huang:2015:HeartGame} the difficulty level, i.e., shaking of the game board, varies depending on the heart rate in order to motivate players to calm down.
Apart from board games, Harley et al. \cite{Harley:2017:TVD} employed tangible objects in virtual reality games in order to improve tactile sensations and experiences beyond those of a classical VR hand-held controller. In one of the tangible objects, i.e., a hollow raccoon toy, the heartbeat was created by an embedded vibration motor.
Magielse et al. \cite{Magielse:2009:OutdoorGame} included the heart rate in an outdoor pervasive game in which the tangible game controllers created beep sounds resembling opponent's heart rate as soon as an opponent player comes in proximity.
In comparison to our work, the discussed papers presented tangible heart rate interfaces in a competitive environment while the focus of our work is on shared user experience in a collaborative and/or cooperative setting.

The second category falls into the art-context where tangible objects make use of heart rate as biofeedback to create physical experiences. For example, Loke et al.  \cite{Loke:2012:HeartArtist} employed the \textit{Bodyweather performance
methodology} and realized a live-art installation in which a participant's breathing and heart rate was mediated through sonification in order to experience those together with an awareness of ``\textit{self, body and the world}''.
Another artistic experience was presented by Nunez-Pacheco et al. \cite{Nunez-Pacheco:2014:ART}, who investigated ``\textit{technology-mediated self-reflection on the body}''. Their installation called ``Eloquent Robes'' makes use of an individual's heart rate to project abstract representations of the measured heart rate on a garment. Through their installation, they found that users who intentionally tried to modify their heart rate had a stronger feeling of self-attachment towards the experience than those just observing with no bodily intervention.
These works pursue similar intentions as our work, that is to attain or reinforce the awareness of one's own heart rate through tangible experiences.
However, while the authors limit their work on individual's experience, the scope of our work is broader and explicitly considers shared experiences of two users in a social context.

The last category focuses on shared and/or remote experiences based on the heart rate of individuals often in a social context to establish an awareness of oneself or others.
Walmink et al. \cite{Walmink:2014:HeartHelmet} studied heart rate displays in a social context in which they displayed the heart rate of a bike rider on the back of cycling helmets. By this means, a group of bike riders have knowledge of the heart rate of others in the group which in turn enables a shared experience and supports engagement within the cycling activity.

Werner et al. \cite{Werner:2008:HeartRings} built a system comprised of two rings capable of measuring the heart rate and sending the data to another corresponding ring. Both rings called “united-pulse” addresses couples who live at a distance and realize a remotely shared experience of the heart rate and creates an ``\textit{artificial corporeality}'' between the couple wearing the rings. They found that minimal tangible output was sufficient to realize a sensual experience.

Another type of shared experience was demonstrated by Hoinkis et al. with ``Herzfassen'' \cite{Hoinkis:2012:Herzfassen}. They utilized a metal bowl filled with water to mediate the heart rate with artificially created waves on the water surface. They build a bass shaker into the bottom of the metal bowl which physically produced the different waves on the water surface. Furthermore, the system re-used the (conductive) handles of the metal bowl for measuring the heart rate of the user. Since the system's purpose is human contact, the metal bowl should be handled between individuals of a group and realizes shared experience as the water surface constantly adapts to a user's heart rate.
The authors noticed that people around an installation tend to play and create a ``chain of arms'' between the bowl handles creating a 'common' heartbeat as part of the shared experience.

The ``Heart Sounds Bench'' of Howell et al. \cite{Howell:2019:HeartBench} addresses the smart city context and aims to ``\textit{invite rest, reflection, and recognition of others' lives in public space}'' which they describe as \textit{life-affirmation}. They integrated a system that records heart sounds through two stethoscopes and speakers to play heart sounds into a red bench. Thereby, people are able to share their heart sounds with others sitting on the bench as well as getting aware of people previously sitting on the bench through their shared heart sounds.
Here, the shared experience helps \textit{recognizing others' lives, feeling connected, and embracing difference with opacity}.

Slovak et al. \cite{Slovak:2012:HRSharing} addressed shared experiences with heart rate measures through a study in which they investigated how people understand their own heart rate and the heart rate of remote users within a so called physio-social space.
They explored the questions with a technology probe and found two distinct categories of effects, that is the heart rate as information and the heart rate as a connection between users. Their work and their implications strongly encourage the use of heart rate communication to support social connectedness or other kinds of social interaction.

While the works discussed so far addresses shared experiences with heart rate interfaces in different social contexts, they explored only non-tangible sensations as heart rate feedback, i.e., visual \cite{Werner:2008:HeartRings, Hoinkis:2012:Herzfassen, Slovak:2012:HRSharing, Walmink:2014:HeartHelmet} or auditory \cite{Howell:2019:HeartBench}.
In contrast to these works, our work includes and focuses on real-time tactile sensations created through the heart rate measures in addition to the visual and auditory feedback channel. 

Finally, Woodward et al. \cite{Woodward:2018:EmoEcho} presented an early prototype called ``EmoEcho'' that aimed to create shared experiences by means of haptic feedback through a tangible cube and a wristband. However, instead of communicating heart rate measures through tactile sensations, their prototype also included touch and motion in order to map the measures to emotions and communicated the emotions between users.

\subsection{Mindfulness and Psychological Wellbeing}
It is not unusual for the inner workings of the human body to be hidden from conscious awareness, allowing users to perform tasks in a more efficient and automated, but often mindless and self unaware manner. For example, we are seldom aware of the cardiac cycle, which is the repeating sequence of events that occur when our hearts beat to circulate blood  through our bodies. The realization of how ``violent'' and fragile the process of our heart's contraction is, is often only realized during videos and images of open-heart surgeries or when a donor's heart (after removal) keeps beating with the help of special devices.  
While there are good reasons to blend away details of the ``cruel'' embodiment for healthy persons, there are also those who argue how bringing the inner workings of our bodies to the foreground for times of reflection can help, for example to recognize and regulate unhealthy behavioral patterns (e.g., \cite{Shusterman2008}). 

Recent developments in computing and interaction design  propagate designs for wellbeing, mindfulness, (Soma)esthetic appreciation \cite{Hook2016},  and mental health \cite{Calvo2016}. To this end, researchers build on foundations and advances in affective computing and user experience research and take inspirations from emerging fields, such as Positive Computing \cite{Calvo2014}, Somaesthetics \cite{Shusterman2008}, and mindfulness based stress reduction (MBSR) \cite{Kabat2003}.

However, mindfulness practices (e.g., meditation techniques to reduce stress  \cite{Kabat2003}) are traditional practices and many related applications and techniques use technology in a very reserved manner, as the presence of technology may distract from focusing one's attention inward. There seems to exist a dichotomy between technology (presence) and naturalness of mindfulness practices, ultimately challenging the appropriation of technology for mindfulness practices.

Zhu et al. \cite{Zhu2017} have recently introduced the term ``Digital Mindfulness'' and provide a scheme consisting of four levels of mindfulness to categorize mindfulness applications. For example, apps that provide multimedia based meditation guidance are in level one. In level two are applications that are personalized. In level three are applications that utilize sensing of user performance and adapt feedback. In level four tools, such as in level one to three disappear, the users becomes one with the ``tool'', and thus, the ``tool'' acts as a gateway to presence. As an analogy of level four Zhu et al. \cite{Zhu2017} propose nature and mindful nature walks, which is a common practice in mindfulness. They present multiple design cases for level four design prototypes.  Interestingly, all the design cases use isomorphic relations (e.g., a glass filled with cold water turns blue) but visual and auditive feedback only, such as movements and color of virtual fish, and sounds of water.

Today, the number of devices and applications measuring and displaying data about ourselves, our bodies, and daily activities has been increasing. Their aim is to provide support and ultimately to increase self-awareness and maintain control over our own physical and mental fitness. 

 While the idea of bringing the inner workings of our bodies to the foreground for times of reflection and mindfulness, and in order to improve wellbeing and happiness seems reasonable and desirable, difficult design challenges exist due to the quality and nature of physiological and behavioral data. 
 As a consequence, the body of work in exploring and transforming traditional self-monitoring practices, techniques to increase self-awareness, and foster self-reflection has been increasing  over the past few years with researchers proposing various ideas to display, utilize, and interact with physiological data.

For example, Gervais et al.\ \cite{Gervais:2016} have proposed a toolkit to expose inner workings of users based on physiological readings. While they use an anthromorphic physical avatar, the physiological signals (e.g., heart rate) and various high level metrics (e.g., workload and arousal) are presented visually (e.g., the avatar has a screen as a face and a light emitting diode embedded in its chest represents its heart).
Roo et al.\ \cite{Roo:2016} have presented ``Inner Garden'', an augmented sandbox, which represents a user's internal state and how the state changes. The feedback is provided through augmented reality (e.g., using a projector) and the sandbox may flourish with plants and flowers growing. While users may use their hands to reshape the sandbox, no haptic feedback is provided about their internal state. Furthermore, virtual avatars and games have been utilized (e.g, \cite{Sonne:2016, shamekhi2015breathe, Qin:2014}) to support  and guide breathing exercises. Fellow researchers have also started to explore the potential of machine learning based approaches, which are challenging but may provide more personalized intelligent solutions (e.g., \cite{Simon18, Ritschel18, rathmann2020towards}).

	\begin{figure}
		\begin{center}
			\includegraphics[width=0.7\columnwidth]{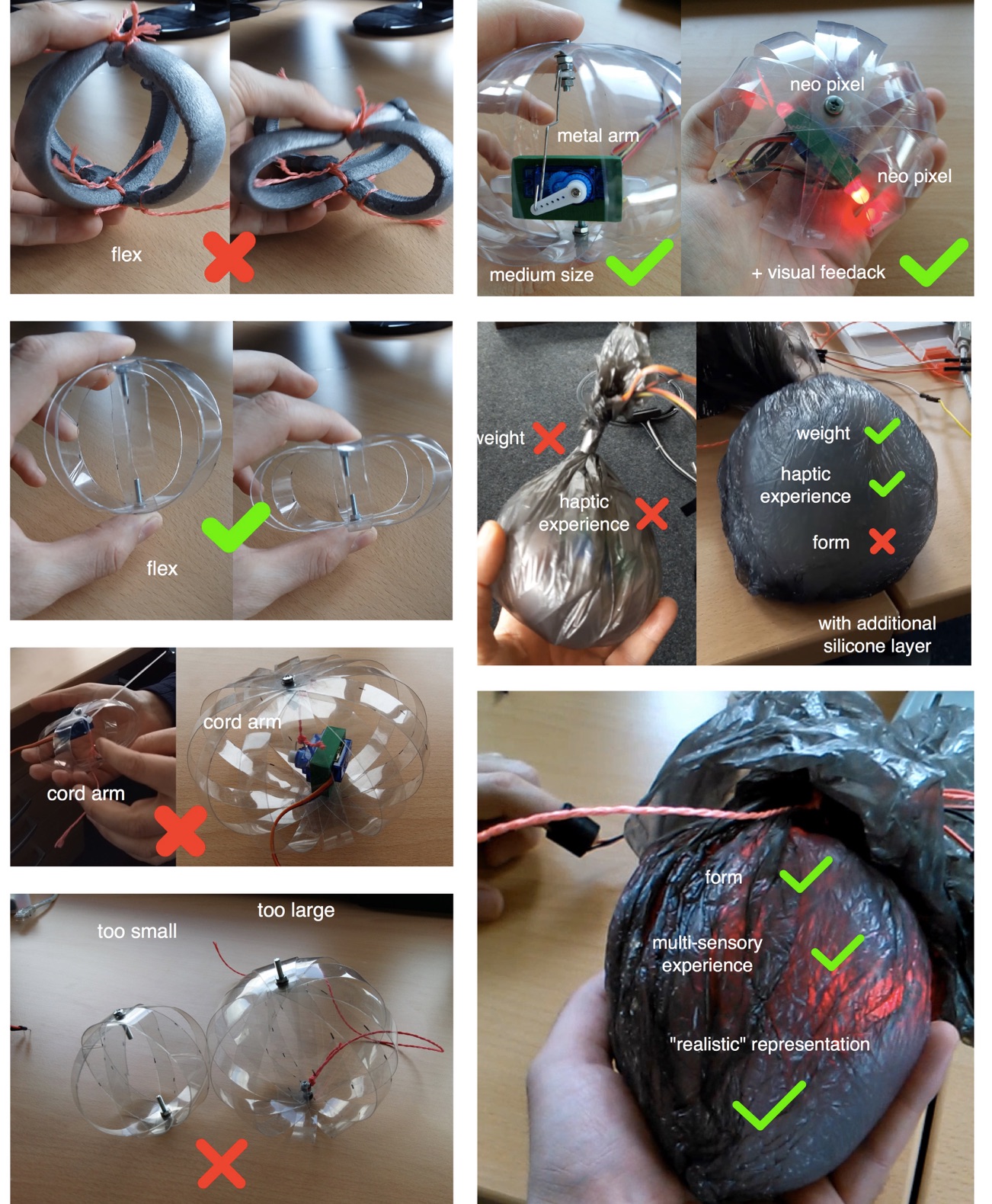}
			\caption{Pictures showing insights into the design/crafting process and some decisions made for the heart design considering materials, form, and feel.}
			\label{fig:design1}
		\end{center}  
	\end{figure}

\section{Design and Implementation of  Tangible Heart Displays}
In Figure \ref*{fig:design1},  we depict stages of the design process and many of the decisions we made in terms of, for example choosing a physical material, form, and dimensions. An off-the-shelf pulse sensor from Adafruit was used to detect heartbeats. To actuate the designs non-continuous servo motors were used.  In order to synchronize sensing and actuation we implemented (hardware) interrupts on an Arduino Uno.

The applied design process was similar to what has been proposed  by Loke and Robertson \cite{Loke:2013} as embodied approaches for movement-based interaction design. For the design process we have also taken an  exploratory approach for the investigations as proposed in design research (e.g.,  \cite{Zimmerman:2007}). That is, we have spent extensive periods of time exploring how the designs feel in our hands through giving physical form to our own physiological data (i.e., heart rate and breathing movements) in different ways and through engaging in phenomenological reflections. Hereby, we have used our own bodies' signals and its movements as material for exploration and developed the necessary ``bodily movement skills'' \cite{Loke:2013}. In our case this meant exploring breathing patterns and movements, and moving our bodies to change our pulse rate.

\subsection{Design Rationality and Iteration}
Experts in somaesthetic practices, such as Shusterman himself, describe episodes when they have been able to clearly feel their beating hearts, including the different places and direction of contradiction \cite{shusterman2012thinking}. However, many people, including the authors, lack the skillful sensitivity for their own heart's behavior. 
While it is clear that the heart is beating non-stop during our lives, its behavior is linked to our feelings through the body-mind connection, which is elaborated extensively in related contemporary work in embodied cognition (e.g., \cite{Chemero:2009, Johnson:2007}). Consequently, stress-inducing factors that influence our feelings also influence our bodies' behavior and thus our heart rate. We chose to use heartbeats since both this ``measure'' is related to meditation training and emotional states associated with stress. 

With the heart design(s) we aimed to replicate the feeling of a beating heart including its weight, form, dimensions, fragility, and movements. Ultimately, the process of exploring different techniques, materials and forms building tangible heart displays helped us establish an expertise for this type of displays, which was helpful in the next steps of building tangible heart displays that could be used in user studies.

To research the user experience of tangible heart displays in a real life setting, we decided to iterate the design and implement prototypes for a multi-user setting, which would allow us to conduct a user study in the field with pairs of users.  Because we utilized a Raspberry Pi for the two new tangible heart displays, we refer to them as \enquote{PiHeart}s.

\begin{figure}[h]
	\begin{center}
		\includegraphics[width=0.7\columnwidth]{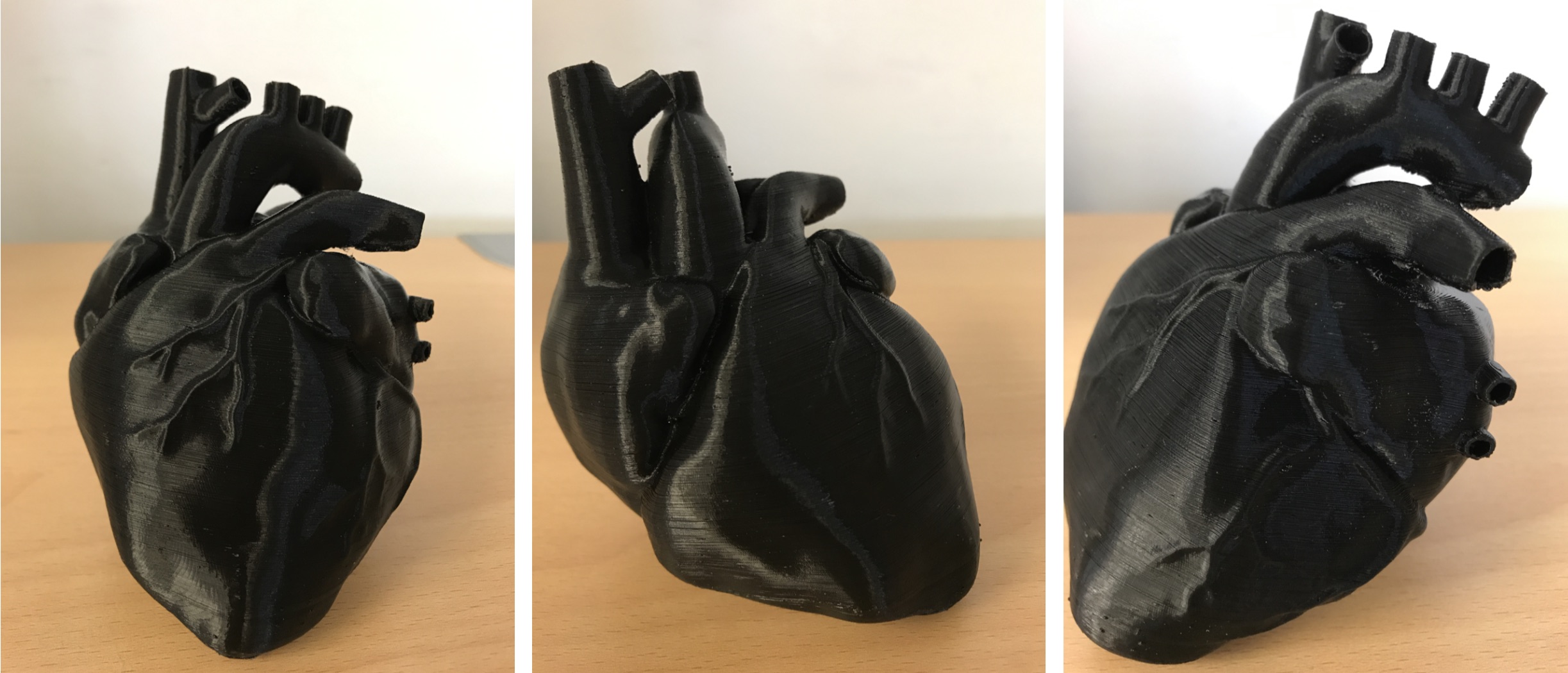}
		\caption{Improving the realism of the shape of the tangible heart design}
		\label{fig:making1}
	\end{center}
\end{figure}

\subsubsection{Materiality and Form of a PiHeart}
For the field study, we decided to improve the materiality and form of our tangible heart displays. To this end, we used a different approach to form  the cover with silicon. We replicated a method, which is often used to produce chocolate figures. First, we printed a realistic shaped heart (see Figure \ref{fig:making1}) and used its shape as template to produce a realistic looking and flexible silicon cover for the PiHearts (see Figure \ref{fig:making1}). One of the reasons why we went back to silicon from latex was  a  strong odor of latex, which was unpleasant.
Ultimately, we used a pouring technique to put the silicon inside the heart shape, which allowed us to create a heart cover that was flexible, and was shaped like a real heart including  the veins on its surface. We had to be careful to choose a very flexible silicon type, which would stay flexible over time. We chose to  use a non-skin color for the silicon, which we believe made the PiHearts appear more aesthetic and playful, and less medical and clinical. 

\begin{figure}[h]
	\begin{center}
		\includegraphics[width=0.7\columnwidth]{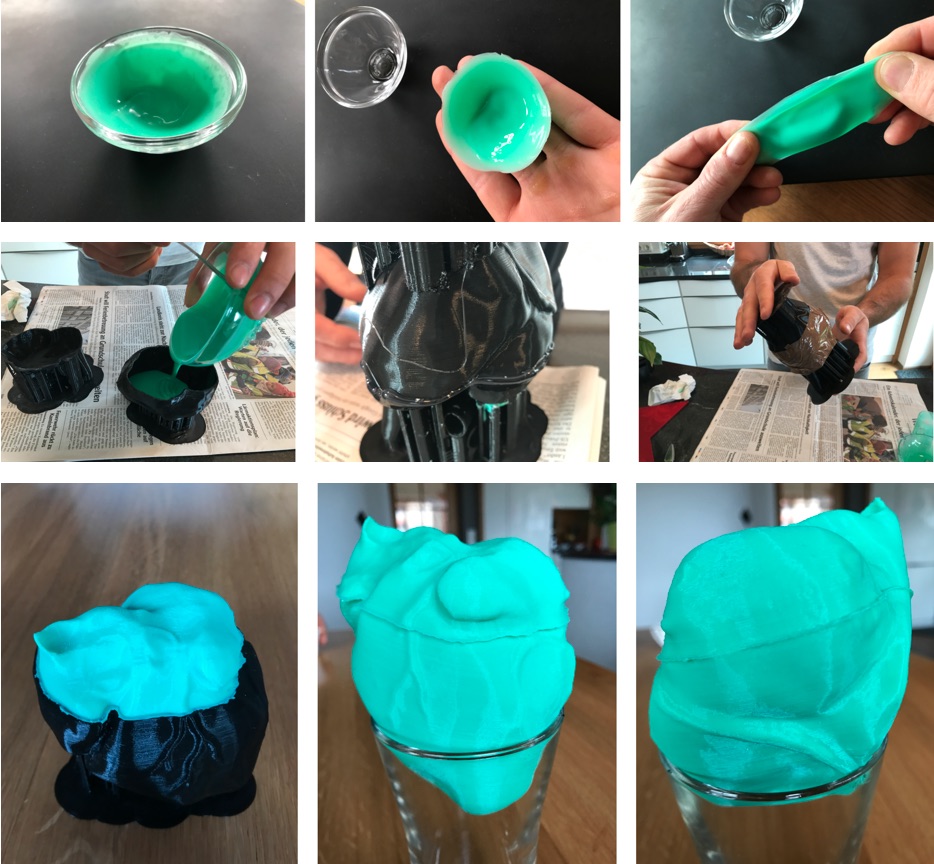}
		\caption{Process of making the PiHeart latex cover.}
		\label{fig:making2}
	\end{center}
\end{figure}

\begin{figure}[h]
	\begin{center}
		\includegraphics[width=0.6\columnwidth]{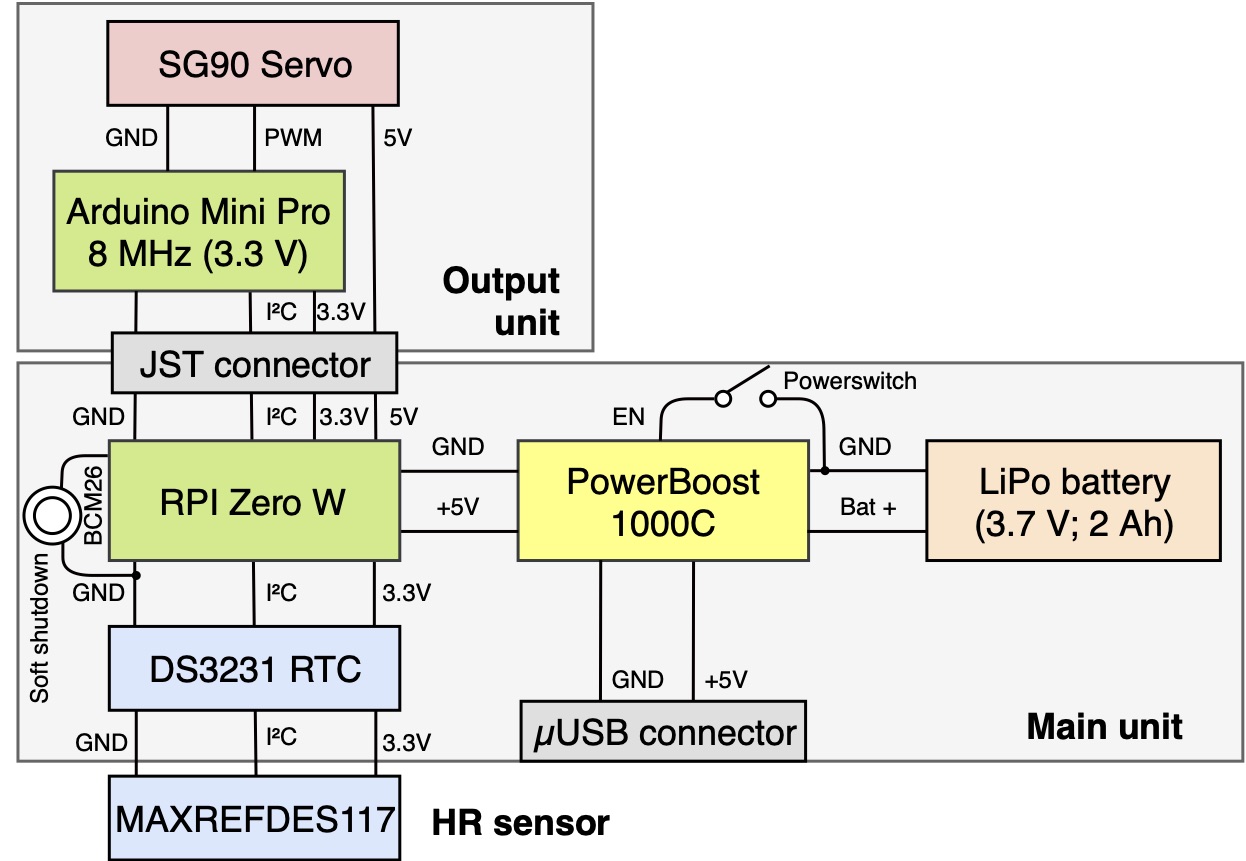}
		\caption{Device connections of the two pairs of main and output units.}
		\label{fig:prototype2}
	\end{center}
\end{figure}

\begin{figure*}[h]
	\begin{center}
		\includegraphics[width=0.8\linewidth]{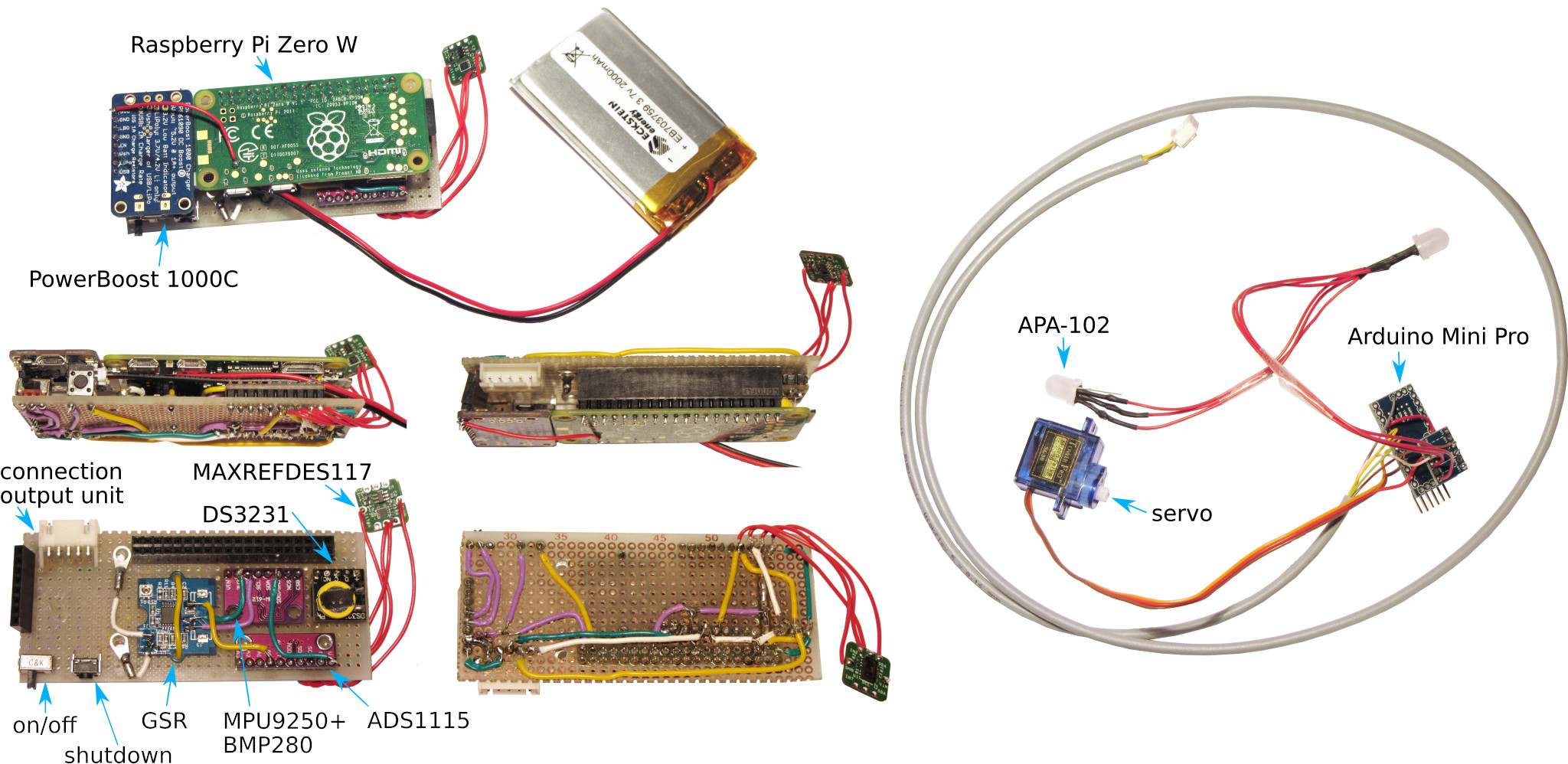}
		\caption{On the left several views of the partly disassembled main unit are visible. The heart rate sensor didn't include the circuit protection and Velcro strip yet. On the right the electronics of the output unit are shown. }
		\label{fig:prototype2_piHeart}
	\end{center}
\end{figure*}

\begin{figure}[h]
	\begin{center}
		\includegraphics[width=0.4\linewidth]{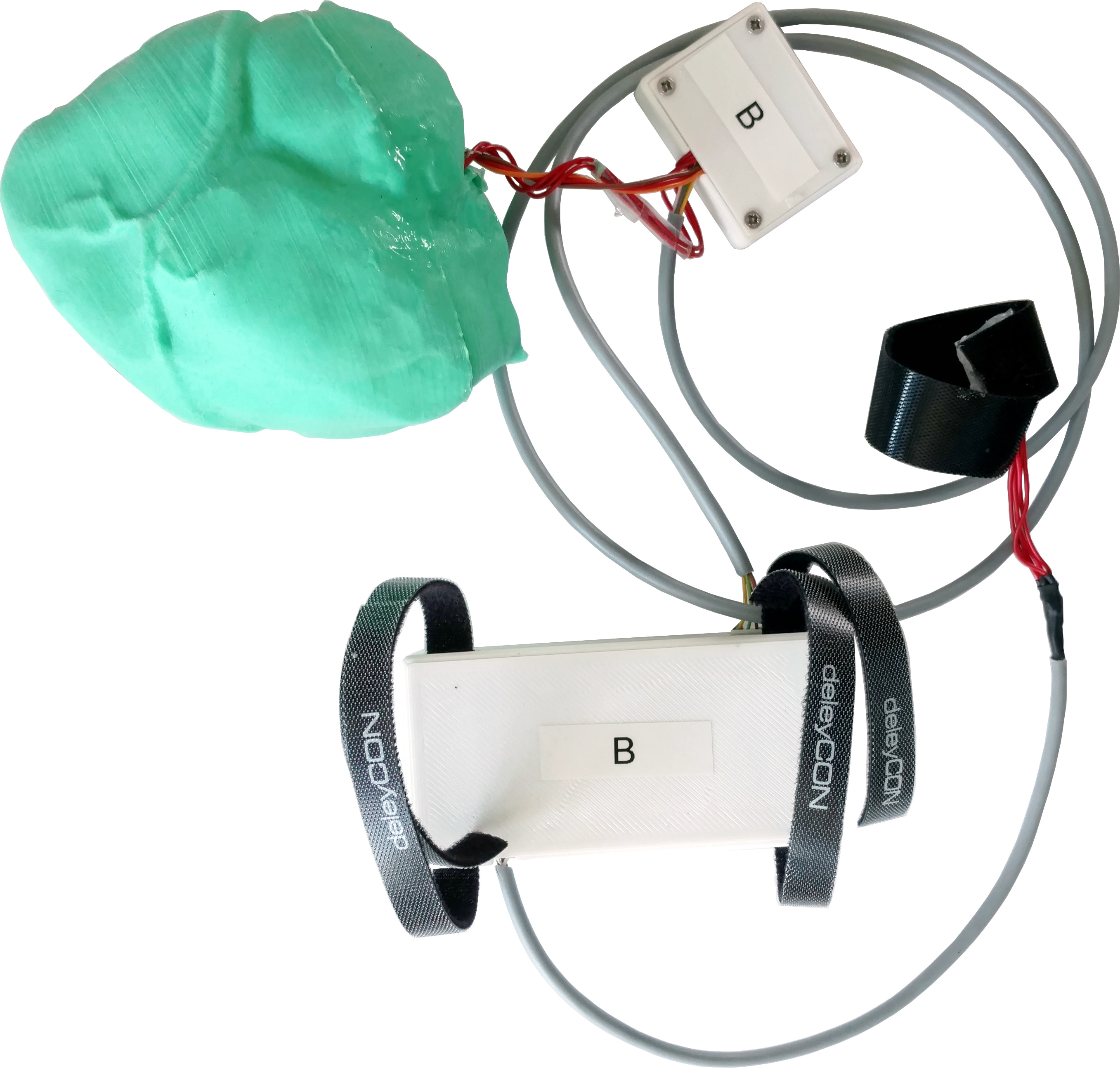}
		\qquad
		\includegraphics[width=0.11\linewidth]{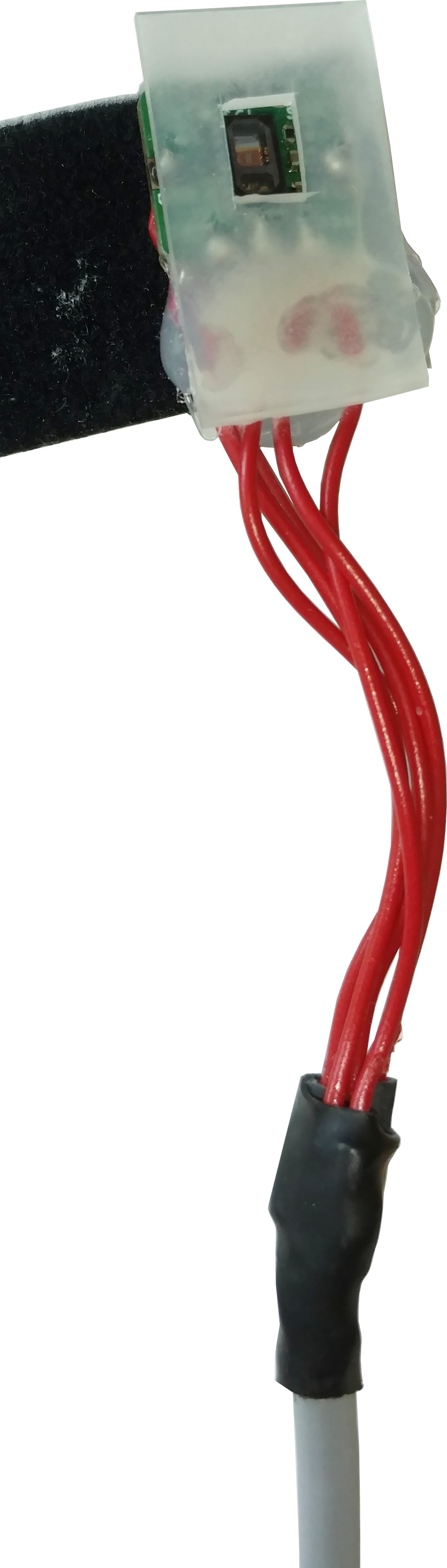}
		\qquad
		\includegraphics[width=0.13\linewidth]{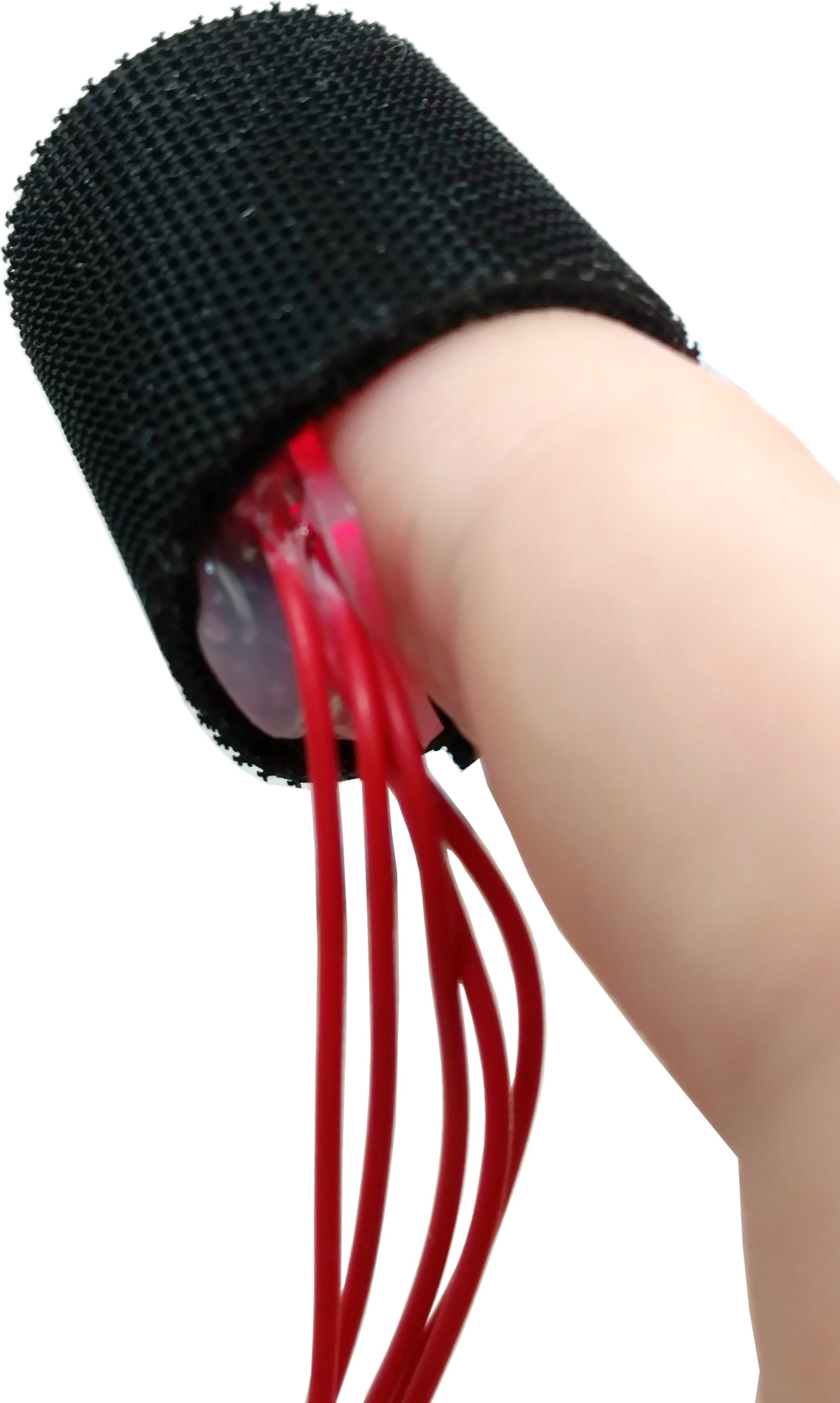}
		\caption{One complete PiHeart prototype including both units enclosed in 3d printed cases and the servomotor inside the silicon heart is shown on the left. On the right the heart rate sensor with Velcro and silicon protection is visible in detail.}
		\label{fig:prototype2_units}
	\end{center}
\end{figure}

\subsection{Hardware}
We crafted two prototypes, whereby a \enquote{PiHeart} prototype consists of two components: a mobile main unit and an output unit. The main unit contains a Raspberry Pi Zero W, all required sensors, and the power supply for all components.  
The output unit includes an Arduino micro-controller that controls two RGB-LEDs and a servomotor. Both units are connected with each other by a plug connection.
Figure~\ref{fig:prototype2} outlines the connections between all components of the prototype. 
Figure~\ref{fig:prototype2_units} shows on the left side a pair of a main unit and an output unit together with their 3D-printed housings and the servomotor enclosed in the silicone heart.
The main unit and the heart rate sensor can be attached to an arm, respectively finger tip, by means of Velcro strips. Figure~\ref{fig:prototype2_piHeart} allows a view of the electronics of the main and output unit.

\paragraph{Main unit}
The main unit contains a LiPo rechargeable battery with a capacity of 2\,Ah connected to an Adafruit PowerBoost 1000C\footnote{\url{https://learn.adafruit.com/adafruit-powerboost-1000c-load-share-usb-charge-boost/overview} (accessed July 10, 2019)} circuit. This circuit contains a TPS61090 boost converter and is able to charge the battery via USB while in use. The circuit delivers a voltage of 5\,V with a maximum current of 2\,A. The power supply to the main unit can be switched on and off with a switch. 
The theoretical maximum current consumption of the main and output unit (1A) allows in our configuration with the capacity of the integrated LiPo battery a runtime of 2 hours.

All other components are connected to the Raspberry Pi Zero W and are partially powered by its 3.3\,V voltage regulator. The Raspberry Pi collects all sensor data and controls all other devices.

One of the two main units includes additional sensors that are included for future research. These are a MPU9250 multi sensor consisting of a three axis accelerometer, gyroscope, magnetometer and a BMP280 air pressure and temperature sensor. Additionally, an ADS1115 16\,bit analog-digital converter is included which is connected to a GSR (galvanic skin response) circuit. The GSR was not used in the studies so that the electrodes were disassembled. Both main units include a DS3231 real time clock (RTC) and a MAXREFDES117\footnote{\url{https://www.maximintegrated.com/en/design/reference-design-center/system-board/6300.html} (accessed July 10, 2019)} circuit with an optical MAX30102 heart rate sensor. We attached a Velcro strip to the sensor breakout board and protected its circuits with a silicon plate with a hole for the optical sensor part. The Velcro strip was responsible to fix the sensor with the finger tip. Figure~\ref{fig:prototype2_units} (rightmost item) depicts the sensor in detail.

We made the Raspberry Pi's I2C interface with a JST plug connection available for external components. The connector additionally provides the voltages 5\,V and 3.3\,V. The 5\,V voltage stems directly from the boost converter and thus can provide higher currents than the 3.3\,V voltage regulator of the Raspberry Pi.

\paragraph{Output unit}
The output unit can optionally be connected with a JST plug connection to the main unit. It is possible to connect several other sensor or actuator devices that can communicate with the main unit via the I2C interface. In this case the output unit consists of an Arduino Mini Pro with a clock of 8\,Mhz powered by the 3.3\,V voltage, an SG90 servomotor and two APA-102 RGB-LEDs which are connected to the Arduino via a levelshifter. One of the two output units does not provide the two RGB-LEDs.  The Arduino acts as I2C slave which receives and executes external commands, in this case from the Raspberry Pi. The servomotor is directly connected to the 5\,V voltage since it draws more power. The required PWM signal can be provided with a voltage of 3.3\,V by the Arduino. For the RGB-LEDs which are also driven with 5\,V the 3.3\,V signals of the Arduino can cause problems so that a levelshifter is used.  

\subsection{Software}
\begin{figure}[h]
	\begin{center}
		\includegraphics[width=0.5\linewidth]{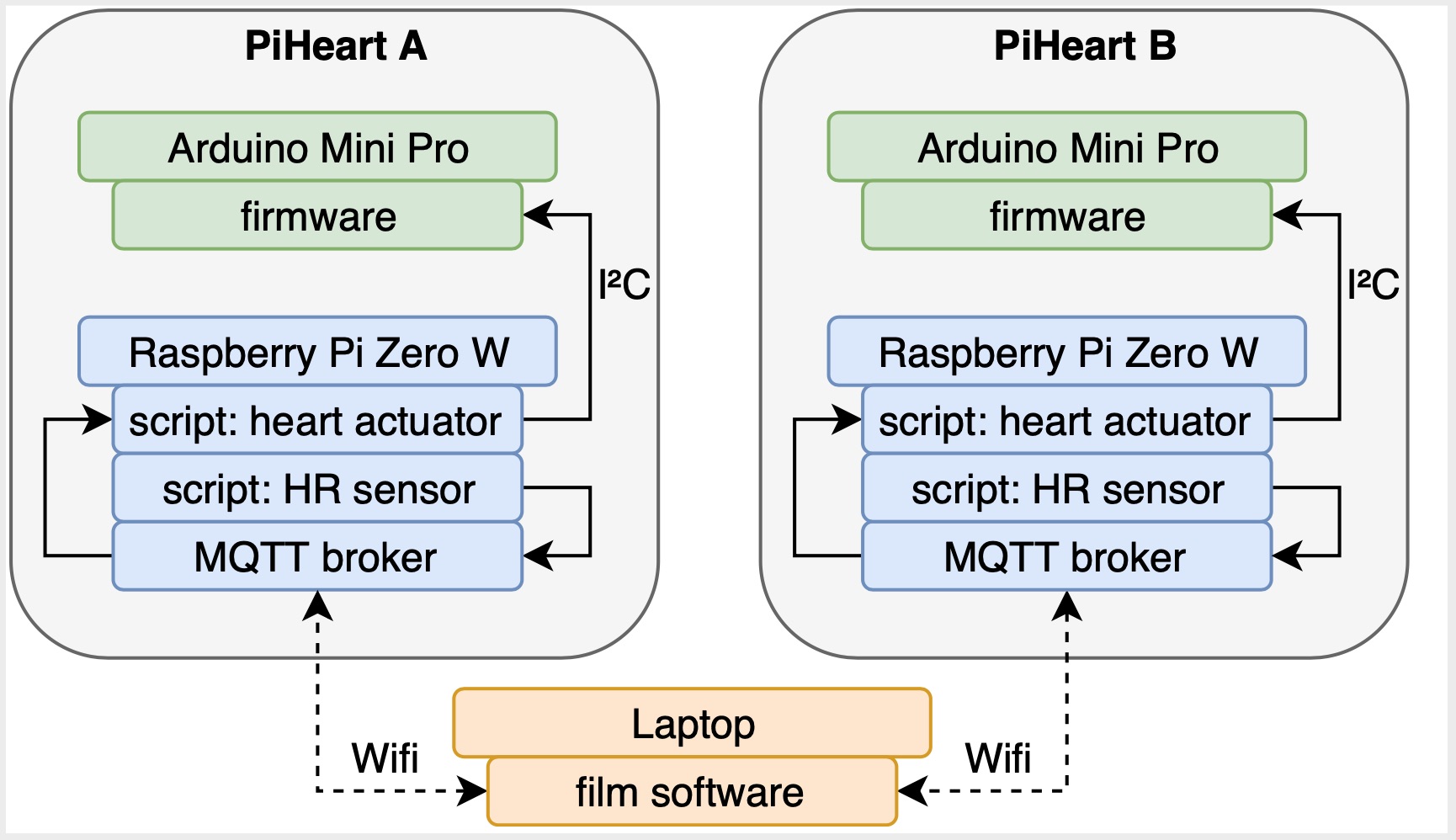}
		\caption{The communication between the devices and the programs.}
		\label{fig:prototype2_comm}
	\end{center}
\end{figure}

An overview of the communication between the software components is sketched in Figure~\ref{fig:prototype2_comm}. In the following, we describe the software of each of the devices.

\paragraph{Arduino}
For the Arduino, we developed a firmware that allows to use it as I2C slave via the I2C bus of the Raspberry. Two libraries for the control of the servomotor and RGB LEDs are used. Nevertheless, the firmware is intended to be minimal so that the software on the Raspberry Pi contains most of the logic. Just one command is implemented which can be sent from the Raspberry. Whenever it is received by the Arduino, a \enquote{heart beat} is conducted. In this case, the axis of the servomotor is set from position zero to 180 and back again while running at full speed. LEDs can optionally shine in a specific color during the movement. Nevertheless, for this study they were disabled.

\paragraph{Raspberry Pi}
\begin{figure}[h]
	\begin{center}
		\includegraphics[width=0.7\linewidth]{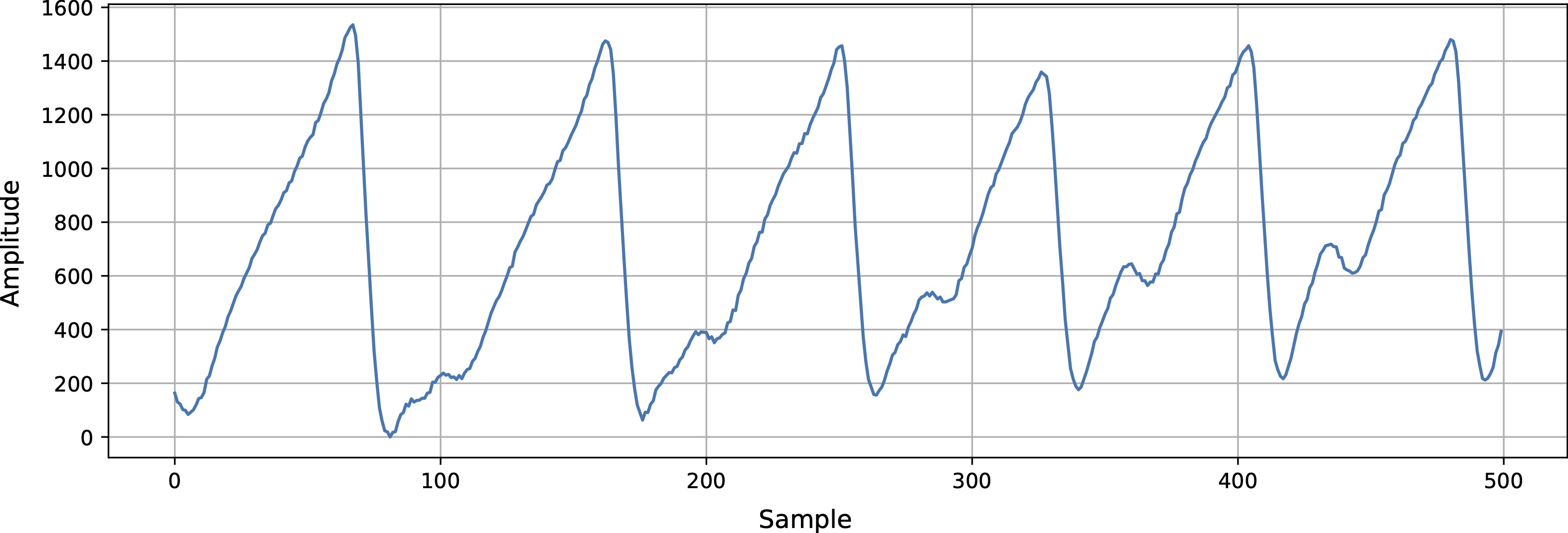}
		\caption{Clean unfiltered BVP sensor data provided by the MAX30102 heart rate sensor at a sample rate of 100\,Hz if positioned correctly and no bigger movements occur.}
		\label{fig:bvp}
	\end{center}
\end{figure}

On the Raspberry, Raspbian Linux was running. It was configured to automatically connect to a specific WiFi-network. Two scripts were developed in Python 3 and communicate with other programs on the network via MQTT protocol. On each Raspberry, a local instance of the MQTT broker \enquote{mosquitto} is running. 

The first script controls the output unit by sending at a specific interval (heart rate) the command to conduct a heart beat to the Arduino via I2C. The heart rate is controlled via MQTT messages.

The second script is more complex. It acquires and calculates the current heart rate of a user from the BVP (blood volume pulse) signal in real time. The BVP signal is provided by the MAX30102 heart rate sensor and is received by the script via I2C. No complex BVP signal artifact compensation is being implemented so that movements should be avoided during the measurements. In our tests the sensor did not provide acceptable signal quality if used at the wrist with several tested sensor configurations (LED brightness, distances) so that we attached it to the finger tip which is a quite common position for this kind of sensor. An example of clean unfiltered data of the sensor is visible in Figure~\ref{fig:bvp}. After the acquisition of the BVP data, the heart rate has to be calculated with the relatively slow single core Raspberry Pi Zero W. In general there are two possibilities. It is possible to calculate the time differences between two peaks which have to be found or a FFT is conducted where the frequency with the highest magnitude is the heart rate. In our experiments we ran into problems with a peak detector module for Python since it required more computational power or the implemented methods were not optimized. The FIFO buffer of the sensor has to be read out in a specific time, if it takes too long due to high CPU load data gets lost which results in wrong heart rate calculations. Thus, we used the FFT approach which required less CPU performance. The MAX30102 was set to provide BVP data at a sample rate of 100\,Hz. The window size of the STFT was set to 30\,seconds (3000 samples) with an overlap of 75\,\% (2250 samples) to be able to calculate the heart rate every 7.5\,seconds. The samples of a window were normalized by their maximum value before applying the FFT. It provides a frequency resolution of 0.033\,Hz which are about 2.0\,bpm. For the calculation of the heart rate just frequencies in the range of 40 to 300\,bpm were considered so that some (movement) artifacts are filtered out. From this range the frequency with the highest absolute value of the real part of the complex numbers is output and provided as MQTT message. The raw BVP signal is additionally sent. The results of this algorithm showed comparable values to a consumer blood pressure meter so that it provided a high enough accuracy for our purposes while providing an easy sensor setup.

\paragraph{Laptop}
A software written in C\# was running on a Windows laptop that controls both PiHeart prototypes, records data and shows the movies. In an SQLite database following data is stored with timestamps: heart rate, the shown movie title, the modality and the raw BVP data. The timestamps created by the laptop software are used for data synchronization as the latencies from data transmission and Wifi connection are usually not very high. In the beginning of each study session the software starts on both Raspberry Pis via SSH connection the two required scripts. The software connects to both MQTT brokers on the Raspberry Pis so that it can communicate with both of them. Since no central MQTT broker is used the laptop can easily be replaced with another device without having to adapt the scripts on the Raspberries. Sending data via network broadcasts was not reliable as some Wifi routers block them.

\begin{figure}[h]
	\begin{center}
		\includegraphics[width=0.8\columnwidth]{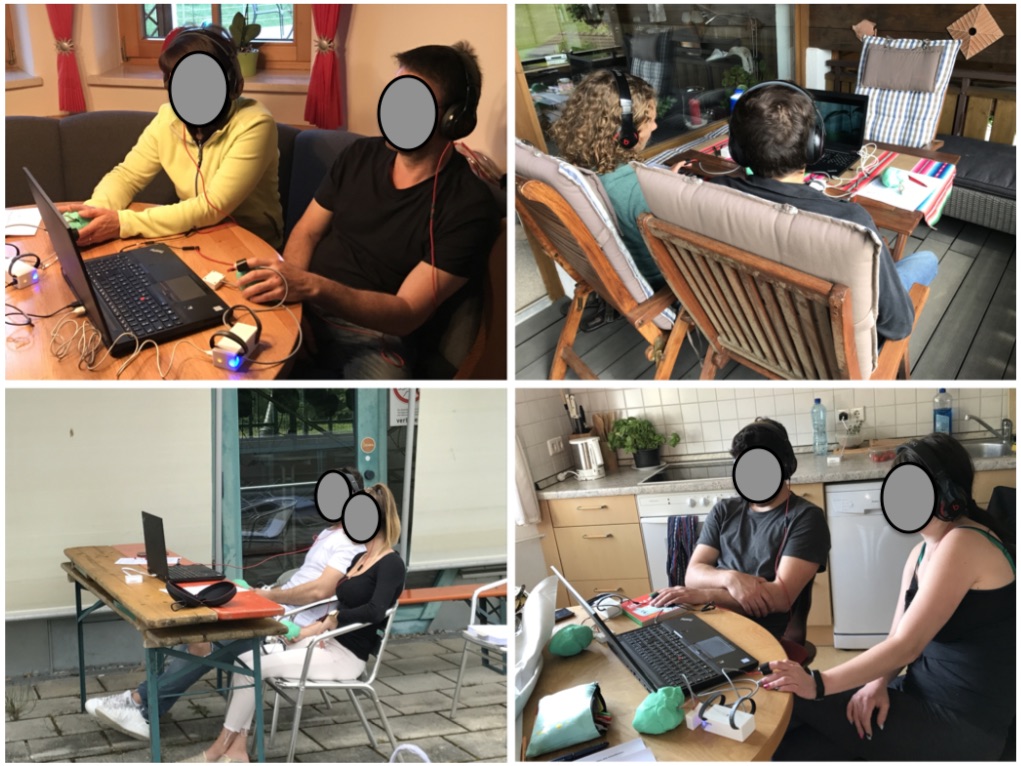}
		\caption{Participants and setup of the field study.}
		\label{fig:field_study}
	\end{center}
\end{figure}

\section{Field Study}

To research how the usage of the PiHearts would influence users' experiences in an everyday social setting we decided to conduct a field study with pairs of participants. Since watching movies is a paradigmatic shared everyday experience, we decided to ask the pairs of participants to watch movies together while using the PiHearts.

We were specifically interested in exploring how the heart displays could potentially change participants' experience by either displaying their heart beats to themselves or to their ``co-viewers''. Thus, the field study aimed at exploring the following research question: \emph{How do the usage of PiHearts potentially change users' experience during a shared movie watching activity when the heart displays feed back (i) users' own heart beats and (ii) when they feed back their ``co-viewers'' heart beats?}
 
Our expectations and believes prior to the study were focussed on experiences associated with how they related to their surrounding world (e.g., immersion and empathy). We believed that by feeding back users their own heart beats with the PiHearts we might be able to strengthen their ``relationship to their bodies'', and thus, increase their mindfulness (e.g., ability to immerse in the moment). We assumed feeding back users their ``co-viewers''' heart beats might in addition increase social awareness (e.g., empathy).  Ultimately, we hoped that the PiHearts would in both conditions improve participants' experiences and potentially foster experiences of resonance (e.g., ``become one with the environment'' through immersion and empathy) and not the opposite (i.e., alienation).

\subsection{Participants and Apparatus}
We recruited 60 participants (30f, 30m) from our own circle of acquaintances for the field study of which 24 reported to be couples in a (romantic) relationship, 4 reported to be friends, and 2 reported to be family members (i.e., sisters or brothers).  Participants' ages  were divers with, for example 14 participants reporting to be between 45 and 65, and 17 between 18 and 24.
The PiHearts, including the laptop, which we described in the previous section were utilized for the field study. Considering the selection movies, we chose  the three movies ``big bunny'', ``overwatch'', and ``for the birds'', which have similar dramatology and similar ratings on internet movie databases such as IMDb.

\subsection{Procedure}
First we conducted a pilot study with two (separate) participants to make sure that our study setup had no major flaws. Afterwards we conducted the study with the 30 pairs of participants within four subsequent weeks. Figure \ref{fig:field_study} depicts exemplary participants and contexts of the field study.  There were three conditions which we studied.  The study was a within-subject study. Consequently, we collected data using a user experience questionnaire  from each pair of subject for each condition. Conditions were watching a movie without any heart beat feedback, watching a movie while holding PiHeart in their hands while PiHeart displayed each participant's own heart beat, and watching a movie while holding PiHeart in their hands and PiHeart displayed their partners heart beats.
The order of the conditions was counter balanced while the order of the movies was fix.  For the field study we decided to choose the three different movies and not one, since we were interested in user experiences, and clearly, watching the same movie a second or third time would strongly influence participants' experience. At the end of the study we conducted a semi-structured asking questions such as ``Please describe the feeling of holding your own heart beats'' and  ``Please describe the feeling of holding the heart beats of the person beside you''. 

In order to measure relevant user experiences we decided to utilize the game experience questionnaire \cite{ijsselsteijn2013game} because this questionnaire measures constructs, such as ``Sensory and Imaginative Immersion'', ``Positive Affect'', ``Negative Affect'', ``Flow'' , which are all relevant in terms of how someone experiences the surrounding world. Moreover, the game experience questionnaire has a social presence module, which measures ``Negative Feelings'', ``Empathy'', and ``Behavioral Involvement'' with all being relevant to a social setting and potentially the notion of social resonance. The game experience questionnaire  was applied after each movie. The analysis of the responses to the open ended questions which were asked at the end (i.e., after participants had watched each movie and reported filled out the user experience questionnaire) was performed by a thematic analysis \cite{braun2006using} in which we identified topics/themes relevant to our research question and sorted the topics/themes in participants' answers by frequency.

\subsection{Results}
\begin{figure}[h]
	\begin{center}
		\includegraphics[width=0.8\columnwidth]{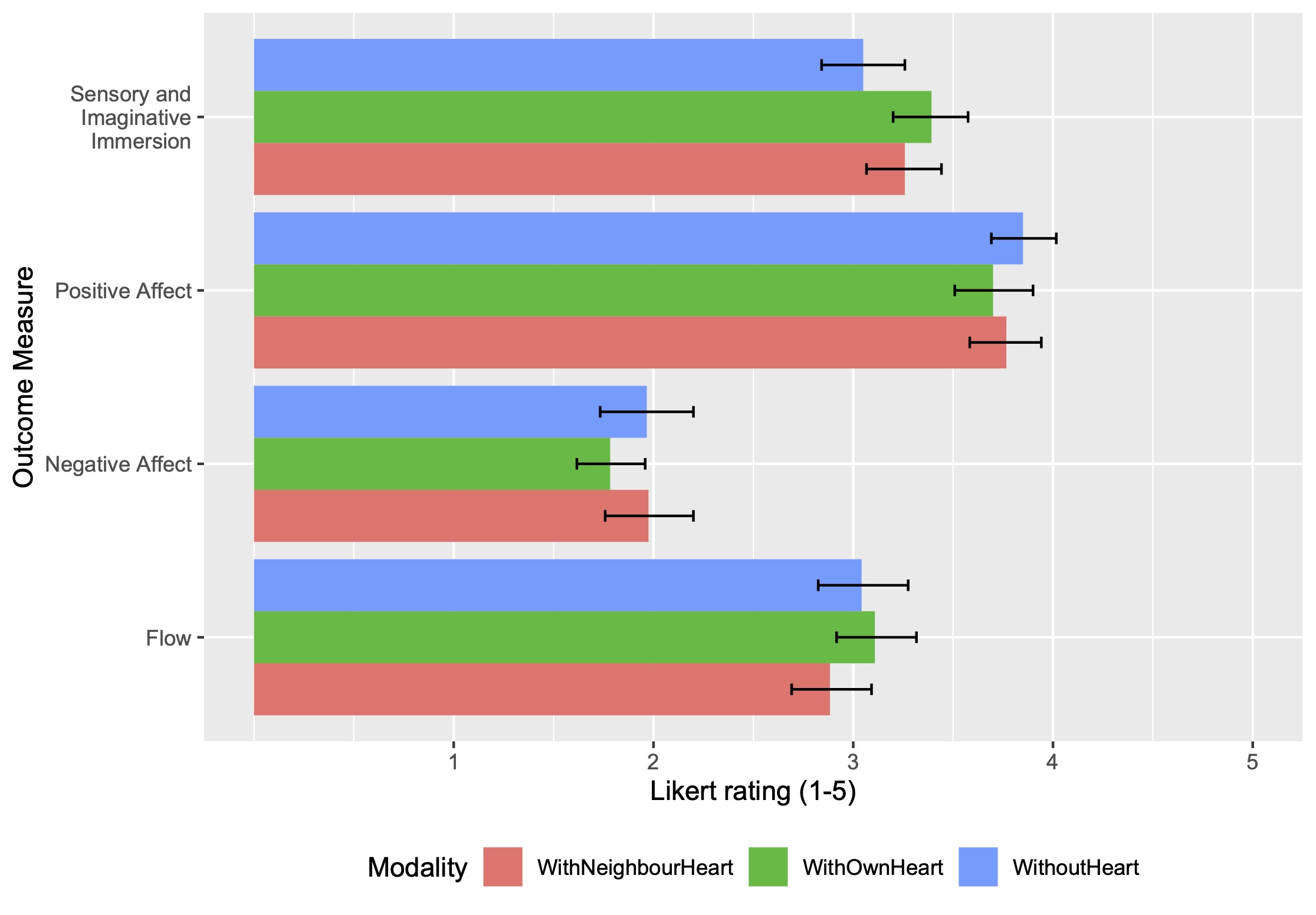}
		\caption{Results of the UX questionnaire}
		\label{fig:results_UX}
	\end{center}
\end{figure}

\begin{figure}[h]
	\begin{center}
		\includegraphics[width=0.8\columnwidth]{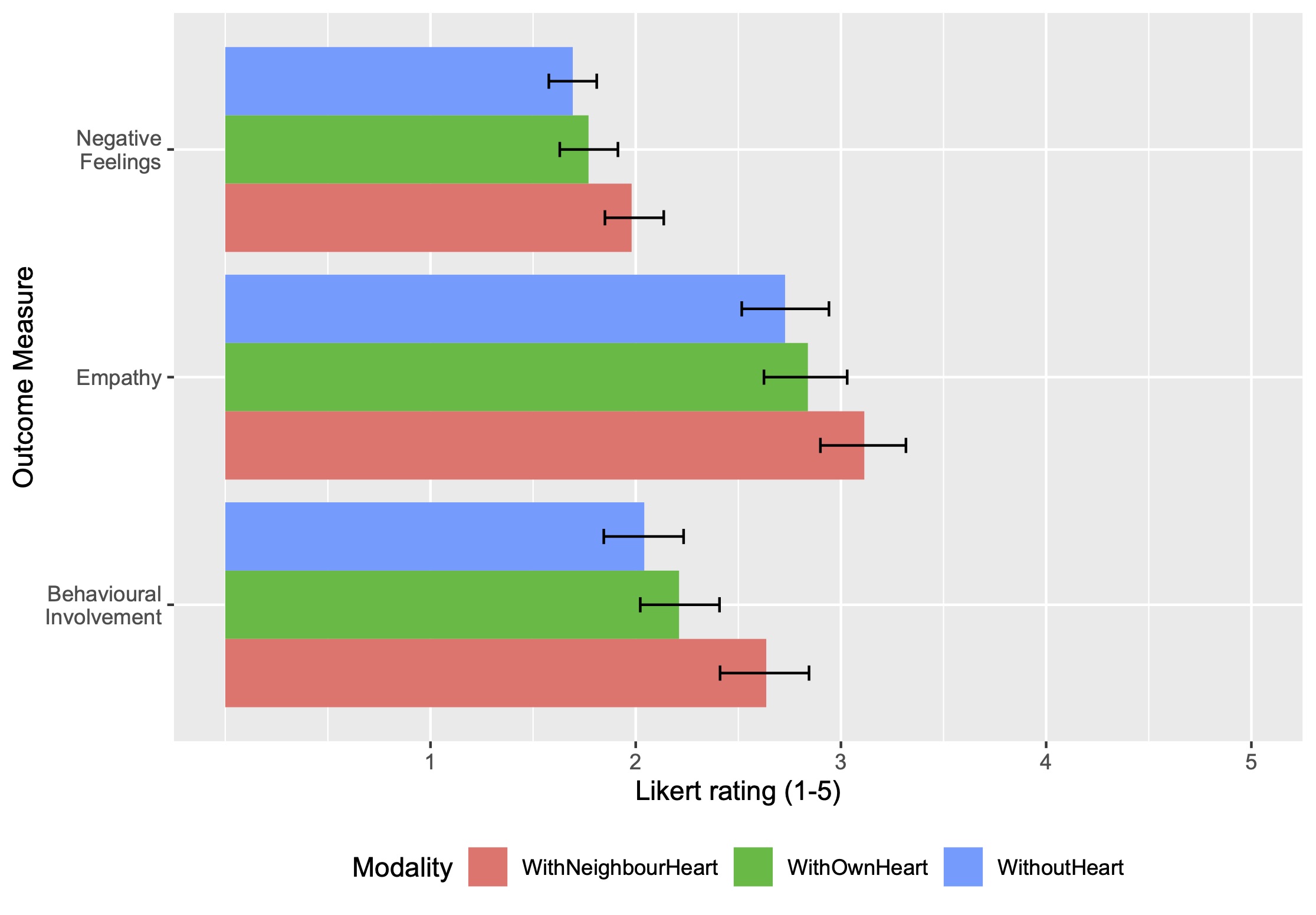}
		\caption{Results of the social presence module}
		\label{fig:results_social_presence}
	\end{center}
\end{figure}

Figure \ref{fig:results_UX} and \ref{fig:results_social_presence} present the results of the UX questionnaire. Figure \ref{fig:results_UX} depicts participants' ratings considering their ``in-movie'' experience and the Figure \ref{fig:results_social_presence} depicts participants' feelings considering social presence.

Table \ref{table:analysis_overview} presents results of a repeated measurement ANOVA and in case of a significant result consequent post-hoc pairwise comparisons (with Bonferroni corrections) for each measured UX construct.  We found a main effect of modality on ``Sensory and Imaginative Immersion''. Pairwise comparisons show that there is a significant difference between the modalities ``WithOwnHeart'' and ``WithoutHeart'', meaning that participant felt significantly higher levels of immersion when they held the PiHeart with their own heart beats being displayed compared to the condition when they didn't have a PiHeart in their hands.
Considering the social presence module, we found a main effect on all three constructs (i.e, ``Empathy'', ``Behavioral Engagement'', and ``Negative Feelings'') with post-hoc pairwise comparisons showing that there are significant differences between the condition when participants used the PiHeart with the neighbors'/partners' heart beats being displayed compared to the other two conditions. When participants watched the movies while holding their neighbors'/partners' heartbeats in their hands, they felt significantly higher levels of social presence. 

\begin{table*}[ht]
	\centering
	\setlength{\tabcolsep}{5pt} 
	\begin{tabular}{ | p{4 cm} | p{1.5cm} | p{2 cm} | p{7 cm} | }
		\hline
		\emph{Dependent variable} & \emph{F-value}& \emph{p-value} & \emph{Post hoc pairwise comparisons (Bonferroni)}\\
		
		\hline
		\hline
		\textbf{Positive Affect} & $F$=1.40 &  p=0.25 &    \\

		\hline
		\hline
		\textbf{Negative Affect}& $F$=1.69 &  p=.19 & \\

		\hline
		\hline
		
		\textbf{Sensory Immersion} &  $F$=4.78 &  \textbf{p=.01 *} & W-N (p=.32); O-N (p=.63); \textbf{O-W (p=.003)}   \\
		
		\hline
		\hline
		\textbf{Flow} &  $F$=2.06 &  p=.13 &  \\	
		
		\hline
		\hline      
		\textbf{Empathy} & $F$=8.55 &    \textbf{p<.001 ***} & \textbf{W-N (p<.001 ***)}; \textbf{O-N (p=.02 *)}; O-W (p=.92)   \\
		
		\hline
		\hline     
		\textbf{Behavioral Engagement} & $F$=16.71 &    \textbf{p<.001 ***} & \textbf{W-N (p<.001 ***)}; \textbf{O-N (p=.0015 **)}; O-W (p=.29)   \\
		
		\hline
		\hline     
		\textbf{Negative Feelings} & $F$=7.53 &    \textbf{p<.001 ***} & \textbf{W-N (p<.001 ***)}; \textbf{O-N (p=.03 *)}; O-W (p=.96) \\
		
		\hline

	\end{tabular}
	\caption{Overview of the statistical tests over all participants, including the overall effect of $modality$  on measures for experience and social presence; and pairwise comparison based on post hoc tests (Signif. codes:  `***' 0.001 `**' 0.01 `*' 0.05   `.' 0.1   ` ' 1). (Abbreviations used for reporting pairwise comparison: WithoutHeart-WithOwnHeart (W-O), WithoutHeart-WithNeighborHeart (W-N), and WithOwnHeart-WithNeighborHeart (O-N))}
	\label{table:analysis_overview}
\end{table*}

\subsection{Analysis of the semistructured interviews}
In the semi-structured interviews, which we conducted at the end of each study session we asked participants a couple of additional questions including the open ended questions of what they liked best and what they liked least.  Most participants (\#17) stated that they liked best the fact that they could feel the heart beats and that the heart rate was not presented in numbers/letters. Followed by 14 participants stating ``feeling the other person's heart beat''.  Three participants explicitly mentioned ``increase of self-awareness considering their heart rate''. 

When we asked participants about what they liked least, most of the participants mentioned one of the movies. Two participants explicitly stated that they least liked being distracted by the heart display.  

The issue that was stated most by participants (\#13) as a potential issue with the usage of a heart display was that one started to get nervous once one felt that the heart beat was rising. When we asked participants what potentials they saw with a tangible heart display the most provided answer (\#12) of participants was related to health and improvement of body consciences followed by couple therapy (\#7).

When we asked participants about what they felt when they held their own heart beat  the most stated answers were  ``it was interesting to feel changes'' and that ``it felt faster than what they would have expected''.  Interestingly the answer that was provided most by participants when we asked them about what they felt when they held their ``partners'' heart beats they said something like ``one feels in competition with each other trying to have the lower heart rate'', which explains why there has been a significant result considering the ``negative feelings''  (often associated with competitive feelings) construct, which is part of the social presence module/questionnaire.

\section{Discussion}

In a fast paced urban life, or as Hartmut Rosa \cite{rosa2019resonance} puts it, due to social acceleration, ``alienation'' may become a health issue with people missing qualitative social interactions. He has argued that because of  social acceleration people seem to desire moments of resonance, which they  search for in  mindfulness apps and doing practices such as yoga, digital detox, etc. Unfortunately, Hartmut Rosa does not provide specifics about how one can design or measure resonance.  However, what we know is that most of our technology-enabled interactions are screen based. For about two decades Hiroshi Ishii \cite{ishii2008tangible} has criticized the experiential qualities of such screen-based human-computer interfaces as unfortunate and that:
\emph{``one can not feel and confirm the virtual existence of digital information through one's hand and body''}. He argued that tangible interface designs will allow us, as humans to experience digital information (with our bodies) in richer ways. 

In this paper, we presented PiHearts, which display in real-time a user's heart beats in a tangible and embodied manner. To explore the user experience of tangible heart displays, we have conducted a field study with 60 participants. We chose to study pairs of users because embodied interaction \cite{Dourish:2001} is essentially about an interweaving of both tangible and social interaction, and because our relations to the surrounding world are defined by both social and tangible ``things'' \cite{rosa2019resonance}.  Thus, our measurement in the field study focused on  participants' ``in-movie'' experience (i.e., how  the movie watching experience was experienced in that moment) and social presence (i.e., participants' social experience) while watching a movie in three different conditions. In two of the conditions participants' experience was augmented with heart beat displays.

\subsection{Experiencing oneself}
We found that when the PiHearts displayed participants own heart beats they reported significantly higher  levels of immersion than when their experience was not augmented with heartbeats. In general, high levels of immersion is arguable a result of being in strong relation to things or persons that one is engaging with.  The utilized questionnaire measures immersion  by measuring interest in the movie story and impressiveness of the experience. Consequently, participants reported to be  more ``interested in the movie's story'' and reported feeling  more impressed when they watched a movie while holding the PiHeart which displayed their own heart beats in their hands.

\subsubsection{Simulated resonance with oneself}
There is a saying that nothing catches one's attention than an image of oneself in the mirror or one's name being called out.  It seems as if integrating one's heart beat into the (movie watching) experience has a similar effect causing participants to experience significantly higher levels of immersion and maybe less lost in other distant thoughts and chaos.
Our results  could mean that there is a chance that feeding back one's heart beat ``simulates social resonance'' and thus reinforces immersion. Put differently, experiencing once rhythmic heartbeat through a tangible display seems to set a person in a state of immersion, and maybe resonance and harmony, because once real heart beat is in synchrony with what is being displayed.  Bennet et al. \cite{Bennett:2015:RBH} have utilized the idea of resonance to explore harmonic interaction with virtual pendulums making use of  subliminal micro-movements. Future work is needed to study the potential of heart beats as design material to serve a similar purpose of harmony and stabilization. Based on our study results, we are hopeful that this is possible in other contexts and designs, and rhythmic patterns of oneself if fed/looped back to users, will similar to the notion of an affective loop (e.g., \cite{hook2009affective, Hook:2016, sundstrom2005exploring})  reinforce immersion and may support experiences of (social) resonance and harmony. 

\subsubsection{Somaesthetics}
We are seldom aware of, for example, the cardiac cycle, which is the repeating sequence of events that occur when our hearts beat to circulate blood  through our bodies. It is not unusual for the inner workings of the human body to be hidden from conscious awareness, allowing users to perform tasks in a more efficient and automated, but often mindless and self unaware manner.  Our work is partially motivated by Somaesthetics (e.g., \cite{shusterman2012thinking}), a ``theory'' and interdisciplinary field that among other things proposes to improve self-awareness through somatic introspection (e.g., body scan exercises). Richard Shusterman who has defined the field of Somaesthetics prefers to use the term ``soma'' instead of the term ``body'', because the later may be associated with undesired prejudice and semantic overload.  With the term soma an emphasis is put on the living body, which in the context of somaesthetic design can be an important  tool in the design process \cite{hook2018designing} itself.

Shusterman advocates improved self-awareness, including sensitivity towards difficult to perceive behaviors, such as one's heart's beating patterns, which are always there but tend to be in the background of a person's awareness.  Through somatic introspections, these unreflected ways of how ones' bodies work are brought to the foreground for example for critical reflection, and thus, to possibly become able to improve behavior. Beyond that, representations about oneself have a fascinating  pull and effect, possibly because most of our senses (as argued by Shusterman) are directed towards the outside and rarely we sense/meet our selves with the exception of using mirrors or viewing media containing representations of our selves, such as selfies. Heart displays can undoubtedly, bring the inner workings of our bodies to the foreground for introspection, without requiring any training or skill, and the tangible nature of  displays may increase the experiential qualities. Ultimately, one may have it easier to become aware of one's behavior and potentially regulate the behavior, and be able to this even in an everyday situation, such as watching a movie.  

\subsection{Experiencing the other}
Results of the field study have also shown a significant effect on social presence when participants' PiHearts displayed their partners' heart beats. We found significant results for all three constructs measured by the presence module. Participants reported significantly higher levels of (i) empathy (e.g., connection to their partners), (ii) negative feelings (e.g., feelings associated with being in a competition), and (iii) behavioral involvement (e.g., believing that they adapted to each other's heart beats).  

\subsubsection{Experiences of resonance and alienation}

Negative feelings were measured by questions associated with feelings of jealousy  or  ``schadenfreude'' (malicious delight), or revengefulness. All these feelings seem associated with participants being competitive. In the interviews, participants mentioned that they wanted their own hearts to beat slower than their partners. Empathy is measured by questions associated with feeling connected with the other, feeling happy when the other is happy, or finding it enjoyable to be with the other. 
Empathy and negative feelings are both sub-constructs of psychological involvement. Ultimately, both feelings were significantly influenced. Dynamics and variation in the relationship between the pairs of users could be one explanation. Another explanation could be that the tangible display reinforced feelings of both, resonance and alienation. Feeling the other person's heartbeat could have provided at times feelings of alienation, when the heart beat pattern of the other was not similar to one-selves. We have experienced that when people experience a tangible heart display, it may cause some irritation (or alienation) in the beginning, because the thing seems ``alive''.  There could be an undesired ``uncanny valley'' effect, which is a phenomenon well known in human-robot interactions, which we did not explore but which may foster feelings of alienation. Future work is needed to uncover details. Until then, it is important that devices, such as the PiHeart need to be used with care because there is potential foster undesired alienation and damage relationships. 
 
Participants also reported significantly higher levels for behavioral involvement when experiencing the other person's heart beats. Behavioral involvement is measured by statements, such as that ones actions depended on the other person's actions, and that  what one did affected the other. Essentially, behavioral involvement is about agency and effectiveness in a reciprocal interaction setting. Hartmut Rosa argues that digitalization is increasing our effectiveness, which is an important aspect for feeling resonance, but digitalization can also increase feelings of ineffectiveness (e.g., when a mobile's battery is down). It seems that participants felt they were effective when using the PiHearts; i.e., being able to adapt to the others heartbeats or feel that the other was adapting to their heartbeats.  

\subsubsection{Social Somaesthetics}
The human body, or soma, is not only a living tool, which has the potential to provide rich self-directed experiences, such as taste and smell. It is a ``face'' that we utilize in social interaction. We perform self-styling, self-fashioning, self-presentation to be perceived as young, cool, beautiful, strong, intelligent, educated, etc.  The social setting and the habit of self-presenting may have cause negative feelings. In the user study, one could argue that the participants tried to self-present themselves by trying to have a slow heart beat, which would be associated with being calm, in control, and healthy.  Indeed, in the user study we asked participants not only to experience themselves but also to experience the other person. One could also argue that we asked them to perform a somaesthetic practice directed towards someone else, such as a masseur providing a massage, which is a category of practices \cite{shusterman2012thinking}.  It is worth noting that providing once heartbeat as a tool or as design material to care and support someone else's body/soma could be a future design concept, which is enabled by ubiquitous computing. 

\subsection{Limitations}
Our research faces some limitations due to the fact that many pairs of participants were in a romantic relationship. It is unclear if similar effects would be achieved if participants' relationship are arbitrary, for example when two people who have never met each other before use the PiHearts. Furthermore, future work is needed to explore if similar effects exists in case ``partners'`` are not co-located but are at remote locations. 

\section{Conclusion}
In this paper we reported on research in designing and evaluating tangible heart displays, which enable the perception of a person's heart beats in an embodied manner.  We presented technical details of the heart displays to allow replicability of the design. We also discussed in detail the results of a field study with 30 pairs of participants evaluating the effect of using the tangible displays during a mundane shared everyday experience (i.e., watching movies together). We found, for example, that participants report significantly higher levels of imaginative and sensory immersion when the tangible heart displays display  user's own heart beats; and participants report significantly higher levels for behavioral engagement when the tangible heart displays display the other person's heart beats.
We have both motivated our research and theoretically grounded it by referring to (i) Somaesthetics as an increasingly interdisciplinary field, which addresses the various uses of the human body and (ii) resonance theory as a modern theory to explain and discuss the effects of using tangible heart displays on social user experiences. We hope that our research will inspire fellow researchers in embodied interaction design and ubiquitous computing to explore how physiological (and rhythmic) data can serve as design material to create new technology-enabled somaesthetic practices and provide experiences of connectedness (e.g., resonance).

\bibliographystyle{unsrt}
\bibliography{references}

\end{document}